\documentclass{ws-ijmpd}

\begin{document}

\markboth{Marco Spaans}
{Stimulated Black Hole Evaporation}

\catchline{}{}{}{}{}
%
%%%%%%%%%%%%%%%%%%%%%%%%%%%%%%%%%%%%%%%%%%%%%%%%%%%%%%%%%%%%%%%%%%%%

\title{STIMULATED BLACK HOLE EVAPORATION}

\author{MARCO SPAANS}

\address{Kapteyn Astronomical Institute, University of Groningen, P.O. Box 800, 9700 AV Groningen, The Netherlands; spaans@astro.rug.nl}

%\author{SECOND AUTHOR}

%\address{Group, Laboratory, Address\\
%City, State ZIP/Zone, Country\\
%second\_author@group.com}

\maketitle

\begin{history}
\received{Day Month Year}
\revised{Day Month Year}
\comby{Managing Editor}
\end{history}

\begin{abstract}

Black holes are extreme expressions of gravity$^1$. Their existence is predicted by Einstein's theory of general relativity and is supported by observations$^{2,3,4}$. Black holes obey quantum mechanics and evaporate spontaneously$^5$. Here it is shown that a mass rate $R_f\sim 3\times 10^{-8} (M_0/M)^{1/2}$ $M_0$ yr$^{-1}$ onto the horizon of a black hole with mass $M$ (in units of solar mass $M_0$) stimulates a black hole into rapid evaporation. Specifically, $\sim 3 M_0$ black holes can emit a large fraction of their mass, and explode, in $M/R_f \sim 3\times 10^7 (M/M_0)^{3/2}$ yr. These stimulated black holes radiate a spectral line power $P \sim 2\times 10^{39} (M_0/M)^{1/2}$ erg s$^{-1}$, at a wavelength $\lambda \sim 3\times 10^5 (M/M_0)$ cm. This prediction can be observationally verified.

\end{abstract}


\section{Quantum Mechanics of Black Holes}

Black holes have been studied for about a century, ever since Schwarzschild found the metric that solves the Einstein equation of general relativity for a point mass$^2$. The Schwarzschild metric is appropriate for a black hole of mass M, zero charge and zero spin1. Characteristic of a black hole are its horizon (a trapped surface) and its singularity. Quantum mechanics and thermodynamics suggest that black holes have a temperature $T \sim 10^{-7} (M_0/M)$ K. This allows them to lose mass through Hawking radiation$^4$.

The context of this work is a black hole and the interaction with its environment. This interaction has quantum mechanical significance because a black hole and its surroundings are not disjoint a priori. Instead, together they establish, through observation, whether a horizon is a net absorber or emitter of mass. Useful quantities to recall below are the Planck length, $l_p \sim 2\times 10^{-33}$ cm, the Planck time, $t_p \sim 5\times 10^{-44}$ s, the Planck mass, $m_p \sim 2\times 10^{-5}$ g and the solar mass, $M_0 \sim 2\times 10^{33}$ g. A black hole's Schwarzschild radius is $R_s \sim 3\times 10^5 (M/M_0)$ cm, $c$ denotes the speed of light, $G$ is the gravitational constant and $h$ is Planck's constant.

The first point to consider is the black hole singularity$^{6,7}$. Its natural habitat is Planckian space-time, characterized by $l_p$, $t_p$ and $m_p$. To explore the quantum properties of singularities, the minimal self-consistent approach is to endow general relativity's smooth space-time manifold with the discrete units $l_p$, $t_p$ and $m_p$, and to pursue their consequences$^{6,8}$. The only dynamical quantity that can be constructed from them is the mass rate $m_p/t_p = c^3/G \sim 6\times 10^{12}$ $M_0$ yr$^{-1}$, which is independent of $h$ and huge. However, in four dimensions the volume $V \sim ct_p l_p^3$ of a Planckian singularity is adjusted to $aV = (a^{1/4}l_p)^4$, with $a = R_s/l_p$, as follows. An observer cannot localize a singularity to better than the linear size of a horizon, anywhere along its surface and at any time, since a singularity can be stretched radially to points on the horizon without an observer able to know. Therefore, the local unit $l_p$ acts through the non-local measure $a^{1/4}l_p$, when one considers the dynamics of a Planckian space-time region enclosed by a horizon. The mass rate $m_p/t_p$ scales as length$^{-2}$ and the horizon-limited quantum rate becomes

\begin{equation}
R_f = (4\pi a^{1/2})^{-1}m_p/t_p \sim 3\times 10^{-8} (M_0/M)^{1/2}\ M_0\ {\rm yr}^{-1},
\end{equation}
for a sphere and proportional to $(hc^{13}/G^5)^{1/4}$. So a black hole endowed with quantum discreteness accommodates mass, by absorption or emission, at a maximum rate $R_f << m_p/t_p$.

The second point to consider is the interplay between $R_f$ and a mass rate onto the horizon, $R$. The mass rates $R$ and $R_f$ can be compared directly when both are measured by the same observer. E.g., one that is in an asymptotically flat region and tracks $M$ as a function of time, $t$, through various astronomical observations. If $dM/dt = R > R_f$ then not all mass should have been added to the black hole, according to the observer. Of course, the flow $R$ is oblivious to this constraint, it obeys space-time curvature and crosses the horizon. Both assertions have to be true, though. They are, in a universe where time is relative, if the black hole eventually loses the excess mass gained by an $R > R_f$ event, provided that the latter event directly causes the required evaporation. This solution implies the existence of a spontaneous and a stimulated evaporation state for a black hole. Stimulated evaporation must proceed at $R_f$, i.e., at a mass rate that reciprocates a maximum absorption trigger with ditto emission. Also, classically the observer detects just black hole mass and has no information on how emitted particles connect to prior accretion. Hence, if activated, stimulated evaporation persists since an observer can only confirm unambiguously that the excess mass acquired by an $R > R_f$ event has been lost if the black hole vanishes.

Define $\psi_+$ to be a conventional black hole and $\psi_-$ to be a stimulated black hole evaporating at $R_f$. The superposition $\psi = (1-\alpha )\psi_+ + \alpha\psi_-$ represents a black hole as a dynamical Planckian space-time region enclosed by a horizon that interacts with an environment. As long as $R-R_f\le 0$ one has $\alpha = 0$, while $R-R_f > 0$ stimulates $\alpha = 0 \rightarrow 1$. A stimulated black hole is not a thermal emitter because power $P = R_fc^2$ scales as $M^{-1/2}$, instead of the usual $M^{-2}$, and is strongly boosted compared to the blackbody rate$^5$. Rather, one has a spectral line with particles of wavelength $\lambda \sim R_s$ across a width $\delta\lambda \sim \lambda /2\pi$, set by horizon geometry as this is the only degree of freedom. The above also pertains to black holes with charge $q$ and angular momentum per unit mass $s$ if the radius $r_+ = M + (M^2 - q^2 - s2)^{1/2}$ is used$^1$. The identity of matter need not be preserved by Planckian space-time, and a traversable wormhole or quantum hair is not part of $\psi$ Indeed, just $l_p$, $t_p$ and $m_p$ are used to describe space-time. Yet, $\alpha = 0 \rightarrow 1$ constitutes an extraction of mass forced by the environment. This enriches the information paradox$^9$.

\section{Implications and Discussion}

Most black holes in the universe comprise stellar origin ones of $\sim 3-30 M_0$. These evaporate in $M/R_f \sim 3\times 10^7 (M/M_0)^{3/2}$ yr if $R > R_f$ holds for brief periods only. But, $R_f \sim 5\times 10^{-9}$ $M_0$ yr$^{-1}$ for a $30 M_0$ black hole and $R > R_f$ is easily achieved, e.g., by an accretion disk$^{8,10}$. The latter is often present and typically radiates at an efficiency $\epsilon \sim 0.1$, with $\epsilon Rc^2$ below the radiation pressure limit of the Eddington luminosity, $L_{edd} \sim 10^{38} (M/M_0)$ erg s$^{-1}$. Also, $M/R_f \sim 5\times 10^9$ yr is cosmologically long and stimulated evaporation is thus a small effect for $\sim 30 M_0$ black holes. Much rarer intermediate mass black holes in globular clusters ($M \sim 10^{2-5} M_0$), and supermassive black holes in centers of galaxies ($M \sim 10^{5-10} M_0$), are also prone to stimulated effects since $R_f$ is even smaller. While $R > R_f$ holds for cosmologically long periods and $M/R_f$ exceeds the age of the universe. Hence, stimulated mass loss is tiny. Primordial black holes of $\sim 10^{15}$ g survive until today. These have $R_f \sim 2$ $M_0$ yr$^{-1}$ and $R > R_f$ appears to be unattainable.

Fortunately, $\sim 3 M_0$ black holes have $R_f \sim 2\times 10^{-8}$ $M_0$ yr$^{-1}$. Rates $R > R_f$ are realisable, but sporadically because $\epsilon R_fc^2 \sim L_{edd}$ and radiation pressure can shut off accretion through a disk$^{10}$. So $R < R_f$ most of the time and $M/R_f \sim 10^8$ yr, which is relatively short. This suggests that a significant fraction of black holes formed at $\sim 3 M_0$ can evaporate, producing a power $P \sim 2\times 10^{39} (M_0/M)^{1/2}$ erg s$^{-1}$. When $M < 10^{-2} M_0$ a stimulated black hole emits photons above the $\sim 10$ MHz cut-off of the earth's ionosphere. In its final year it converts $M \sim 10^{-5} M_0$ (so $\sim 10^{49}$ erg) into $>10$ GHz photons. The last second produces $P > 10^{44}$ erg s$^{-1}$ at $>10^{15}$ Hz, comparable to an entire galaxy. These explosive final stages should provide valuable probes of the standard model of particle physics.

%\begin{equation}
%\mu(n, t) = \frac{\sum^\infty_{i=1} 1(d_i < t, N(d_i) 
%= n)}{\int^t_{\sigma=0} 1(N(\sigma) = n)d\sigma}\,.
%\label{eqn1}
%\end{equation}
%Equations should be referred to in abbreviated form,
%e.g.~``Eq.~(\ref{eqn1})'' or ``(2)''. In multiple-line
%equations, the number should be given on the last line.

%\section*{Acknowledgments}
%
%This section should come before the References. Dedications and funding 
%information may also be included here.

%\begin{thebibliography}{000} %for 3 digits
%\begin{thebibliography}{00}  %for 2 digits

\end{document}